\documentclass[12pt]{iopart}

\usepackage{iopams}
\usepackage{bbm}
\usepackage{bbold}
\def\W{\wedge}
\def\D{\mbox{d}}
\def\slash#1{\, /\kern-0.6em{#1}}
\def\genf#1#2{\buildrel#2\over{\mathfrak #1}}
\def\gend{\mathbf d}
\def\lie{\hat{\mathcal L}}

\begin{document}

\title{Generalized forms and vector fields}     
\author{Saikat Chatterjee, Amitabha Lahiri and Partha Guha}
\address{S. N. Bose National Centre for Basic Sciences, \\ 
Block JD, Sector III, Salt Lake, Calcutta 700 098, INDIA\\ 
saikat, amitabha, partha, @boson.bose.res.in}

\begin{abstract}
The generalized vector is defined on an $n$ dimensional manifold.
Interior product, Lie derivative acting on generalized $p$-forms,
$-1\le p\le n$ are introduced. Generalized commutator of two
generalized vectors are defined. Adding a correction term to
Cartan's formula the generalized Lie derivative's action on a
generalized vector field is defined. We explore various identities
of the generalized Lie derivative with respect to generalized
vector fields, and discuss an application.
\end{abstract}

\pacs{02.40.Hw, 45.10.Na, 45.20.Jj}

%%%%%%%%%%%%%%%%%%%%%%%%%
\section{\label{intro}Introduction}
%%%%%%%%%%%%%%%%%%%%%%%%%
The idea of a $-1$-form, i.e., a form of negative degree, was first
introduced by Sparling~\cite{Sparling:avrc97, Perjes:esi98} during
an attempt to associate an abstract twistor space to any real
analytic spacetime obeying vacuum Einstein's equations. Sparling
obtained these forms from the equation of the twistor surfaces
without torsion. There he assumed the existence of a $-1$-form.

Nurowski and Robinson~\cite{nurob:cqgl01, nurob:cqg02} took this
idea and used it to develop a structure of generalized differential
forms. They studied Cartan's structure equation, Hodge star
operator, codifferential and Laplacian operators on the space of
generalized differential forms. They found a number of physical
applications to mechanical and physical field theories like
generalized form of Hamiltonian systems, scalar fields, Maxwell and
Yang-Mills fields and Einstein's vacuum field equations.  Guo et
al.~\cite{Guo:2002yc} later found that Chern Simons theories can be
related to gravity using the language of generalized forms.
Robinson~\cite{Robinson:jmp03} extended the algebra and calculus of
generalized differential forms to type $N$, where $N$ is the number
of independent $-1$-form fields in the structure. $N=1$ represents
the ordered pair representation of~\cite{nurob:cqgl01,
nurob:cqg02}. We also work in $N=1$ case.  Action of ordinary
vector fields on generalized forms are discussed by Nurowski and
Robinson~\cite{nurob:cqg02}, i.e. interior product, Lie derivative
of generalized forms with respect to ordinary vector fields. In
this paper we introduce the generalized vector field $V$ as an
ordered pair of an ordinary vector field $v_1$ and ordinary scalar
field $v_0$, $V=(v_1,v_0)$.

In \S\ref{genf} we briefly discuss generalized forms, their
products and exterior derivative developed by Sparling and
Nurowski-Robinson. In \S\ref{genv} we define the interior
product of a generalized $p$-form with a generalized vector field
as a mapping from generalized $p$-forms to generalized $p-1$-forms
and various identities are discussed. In \S\ref{lieform} we
define the generalized Lie derivative of a generalized $p$-form
using Cartan's formula. It follows that a generalized bracket of
two generalized vector fields can be defined and generalized vector
fields form a Lie algebra with this generalized bracket. But the
Lie derivative of a vector field cannot be defined using Cartan's
formula. So in \S\ref{lievector} we add a correction term
with Cartan's formula and define a new and improved generalized Lie
derivative. This allows us to define the Lie derivative of a
generalized vector field. The Lie derivative satisfies Leibniz rule
and linearity, while the definition of generalized commutator
remains the same. In \S\ref{altvect} we give an alternative
construction of the generalized vector field assuming the existence
of a special kind of vector field. Its properties are discussed and
the results compared with the relations obtained earlier. Finally,
an example is given in \S\ref{relpart}, where we discuss the
Hamiltonian of the relativistic free particle in the language of
generalized forms and generalized vector fields.

%%%%%%%%%%%%%%%%%%%%%%%%%
\section{\label{genf}Generalized $p$-forms}
%%%%%%%%%%%%%%%%%%%%%%%%%
Let $M$ be a smooth $n$-dimensional manifold, $C^{\infty}(M)$ the
ring of real-valued smooth functions, ${\cal X}(M)$ the Lie algebra
of vector fields and $\Omega^p(M)$ the $C^{\infty}(M)$-module of
differential $p$-forms, $ 1 \leq p \leq n$.

A generalized $p$-form $\genf{a}{p}$ on $M$ is
defined~\cite{nurob:cqgl01, nurob:cqg02} as an ordered pair of an
ordinary $p$-form $\alpha_p$ and an ordinary $p+1$-form
$\alpha_{p+1}$,
\begin{equation}
\genf a p \, = 
(\alpha_p,\,\alpha_{p+1})\,. \qquad\qquad -1\leq p\leq n\,
\label{genf.def}
\end{equation}
Since there is no such thing as an ordinary $-1$-form or an
ordinary $n+1$-form (Sparling's $-1$-form is by no means ordinary,
as we shall discuss later), 
\begin{equation}
\genf a{-1} = (0, \alpha_0)\,,\qquad \genf a{n} = (\alpha_n, 0)\,. 
%\label{}
\end{equation}
We will denote the space of generalized $p$-form fields on $M$ by
$\Omega^p_G(M)$.

The wedge product of a generalized $p$-form
$\genf a p\,=(\alpha_p,\,\alpha_{p+1})$ and a generalized $q$-form
$\genf b q \,=(\beta_q,\,\beta_{q+1})$ is a map
$\wedge:\,\Omega^p_G(M)\times \Omega^q_G(M) \to \Omega^{p+q}_G(M)$
defined as~\cite{nurob:cqgl01}
\begin{eqnarray}
\genf a p \W \genf b q = (\alpha_p\beta_q,\, \alpha_p\beta_{q+1} +
(-1)^q \alpha_{p+1} \beta_q).
\label{genf.wedge}
\end{eqnarray}
Here and below, we write $\alpha_p\beta_q$ for the (ordinary) wedge
product of ordinary forms $\alpha_p$ and $\beta_q$\,, without the
$\wedge$ symbol, so as to keep the equations relatively
uncluttered. We will use the $\wedge$ symbol only to denote the
wedge product of generalized forms. Clearly, the wedge product as
defined above satisfies
\begin{equation}
\genf{a}{p} \W \genf{b}{q}\, = 
(-1)^{pq} \genf{b}{q} \W  \genf{a}{p}\,,
\label{genf.trans}
\end{equation}
exactly like the wedge product of ordinary forms.

The generalized exterior derivative $\gend :
\Omega^p_G\rightarrow \Omega^{p+1}_G$ is defined as
\begin{equation}
\gend \genf{a}{p}\, = 
(\D\alpha_p+(-1)^{p+1}k\alpha_{p+1},\,\D\alpha_{p+1}).
\label{genf.extd}
\end{equation}
Here $\D$ is the ordinary exterior derivative acting on ordinary
$p$-forms and $k$ is a nonvanishing constant.  It is easy to check
that the generalized exterior derivative satisfies the Leibniz rule
on generalized forms,
\begin{eqnarray}
\gend ( \genf{a}{p} \W  \genf{b}{q}) &=& 
(\gend  \genf{a}{p})\W  \genf{b}{q} + 
(-1)^{p}  \genf{a}{p} \W (\gend  \genf{b}{q})\,.
\label{genf.Leibniz}
\end{eqnarray}

We make a small digression here to note that there is an
alternative definition of the generalized $p$-formas we mentioned
at the beginning. In this we start by assuming, following
Sparling\cite{Sparling:avrc97}, the existence of a $(-1)$-form
field, i.e., a form of degree $-1\,.$ Denoting this form by
$\zeta\,,$ and using $\D^2 = 0\,,$ we find that we should have
$\D\zeta = k\,,$ with $k$ a constant. Then a generalized $p$-form
is defined in terms of an ordinary $p$-form $\alpha_p$ and an
ordinary $(p+1)$-form $\alpha_{p+1}$ as
\begin{equation}
\genf{a}{p}\, = \alpha_p + \alpha_{p+1}\W\zeta\,.
\label{genf.-1}
\end{equation}
Note that $\zeta$ is not a differential form in the usual sense,
i.e., given an ordinary one-form $\alpha_1$, the wedge product
$\alpha_1\W\zeta$ is not a function on the manifold. However, if we
nevertheless treat the object of Eq.~(\ref{genf.-1}) as an ordinary
$p$-form, we find using $\zeta\W\zeta = 0\,$ that the wedge product
of two such forms is
\begin{eqnarray}
\genf{a}{p}\W \genf{b}{q} &=& \alpha_p\beta_q +
\left(\alpha_p\beta_{q+1} + (-1)^q \alpha_{p+1}
\beta_q\right)\W\zeta\,, 
\label{genf.-1wedge}
\end{eqnarray}
while the exterior derivative of such a $p$-form works out to be
\begin{equation}
\gend \genf{a}{p}\, = 
\D\alpha_p+(-1)^{p+1}k\alpha_{p+1} + \D\alpha_{p+1}\W\zeta\,.
%\label{}
\end{equation}
The two definitions in Eq.s~(\ref{genf.def}) and (\ref{genf.-1})
are clearly equivalent. We will employ the first definition in all
calculations.

%%%%%%%%%%%%%%%%%%%%%%%%%
\section{\label{genv}Generalized vectors and contraction} 
%%%%%%%%%%%%%%%%%%%%%%%%%
In order to go beyond forms and define generalized tensor fields,
we will need to define a generalized vector field. We could define
a generalized vector as an ordered pair of a vector and a bivector,
dual to a generalized one-form, referring either to the ring of
real functions, or to the ring of generalized zero-forms. Both
these definitions quickly run into problems. The interior product
with generalized forms is ill-defined in one case, and fails to
satisfy Leibniz rule in the other. In this paper we propose a new
definition of the generalized vector field. 

%%%%%%%%%%%%%%%%%%%%%%%%%
\subsection{Generalized vector fields}
%%%%%%%%%%%%%%%%%%%%%%%%%
We define a generalized vector field $V$ to be an ordered pair of
an ordinary vector field $v_1$ and an ordinary scalar field $v_0$,
\begin{equation}
V := (v_1, v_0)\,, 
\qquad v_{1}\in {\cal X}(M)\,,\, v_{0}\in C^{\infty}(M).  
\label{genv.def}  
\end{equation}
Clearly, the submodule $v_{0}=0$ of generalized vector fields can
be identified with the module of ordinary vector fields on the
manifold. We will write ${\cal X}_G(M)$ for the space of
generalized vector fields on $M\,.$ On this space, ordinary scalar
multiplication is defined by
\begin{equation}
\lambda V = (\lambda v_1, \lambda v_0)\,,
\label{genv.scm}
\end{equation}
as expected. On the other hand, we can define {\em generalized
scalar multiplication}, by a generalized zero-form $\genf a0\, =
(\alpha_0, \alpha_1)$, by
\begin{equation}
\genf a0 V = (\alpha_0 v_1, \alpha_0 v_0 + i_{v_1}\alpha_1)\, \in
{\cal X}_G(M) \,,
\label{genv.gscm}
\end{equation}
where $i$ is the usual contraction operator. Note that on the
submodule $v_0 = 0\,,$ i.e., on ordinary vector fields, the
generalized scalar multiplication is different from the ordinary
scalar multiplication given in Eq.~(\ref{genv.scm}). The
generalized scalar multiplication is linear, and satisfies
\begin{equation}
\genf a0 (\genf b0 V) = (\genf a0 \W \genf b0) V\,.
\label{genv.gscmring}
\end{equation}
%

%%%%%%%%%%%%%%%%%%%%%%%%%
\subsection{Generalized contraction}
%%%%%%%%%%%%%%%%%%%%%%%%%
Next we define the generalized contraction, or interior product
$I_V$, with respect to a generalized vector $V = (v_1, v_0)$, as a
map from generalized $p$-forms to generalized $(p-1)$-forms.  We
will define this mapping in such a way that, given a generalized
$p$-form $\genf ap = (\alpha_p, \alpha_{p+1})\,,$ setting $v_0 = 0$
and $\alpha_{p+1} = 0$ gives the contraction formula of an ordinary
vector with an ordinary $p$-form. Then we can define the most
general contraction formula $I_V:\Omega^{p}_{G} \rightarrow
\Omega^{p-1}_{G}$ as
\begin{equation}
I_V \genf{a}{p}\, = 
(i_{v_{1}}\alpha_p\,, \lambda(p)\,i_{v_{1}}\alpha_{p+1} +
\tau(p)\,v_{0}\alpha_p)\,,
\label{genv.geni}
\end{equation}
where $\lambda(p)$ and $\tau(p)$ are unknown functions. 

Now we impose Leibniz rule on the contraction formula, so that
$I_V$ is a $\wedge$-anti-derivation,
\begin{equation}
I_V (\genf{a}{p} \W \genf{b}{q}) = 
(I_V \genf{a}{p})\W \genf{b}{q} + 
(-1)^{p}\genf{a}{p} \W (I_V \genf{b}{q})\,.
\label{genv.antider}
\end{equation}
We find from this condition, using Eqs.~(\ref{genf.wedge}) and
(\ref{genv.geni}), that 
\begin{equation}
\lambda(p) = 1\,,
\end{equation}
and
\begin{eqnarray}
(-1)^q\tau(p) + (-1)^p\tau(q) &=& \tau(p+q)\,.
\label{genv.taulambda}
\end{eqnarray}
The non-trivial solution of this is $\tau(p) = p(-1)^{p-1}\,,$ up
to an arbitrary constant, which we will set to unity.

Thus the formula for contraction of a generalized vector field with
a generalized $p$-form is
\begin{eqnarray}
I_V \genf{a}{p}\, = 
(i_{v_{1}}\alpha_p\,, i_{v_{1}}\alpha_{p+1} +
p(-1)^{p-1}v_{0}\alpha_p)\,.
\label{genv.contract}
\end{eqnarray}
The interior product is linear under ordinary scalar multiplication,
\begin{equation}
I_{V+\mu W}= I_V +\mu I_W,  
\label{genv.linear}
\end{equation}
where $\mu$ is an ordinary scalar field, but it fails to
satisfy linearity under generalized scalar multiplication,
\begin{eqnarray}
I_{\genf a0 V} \genf bq &\neq& \genf a0 (I_V \genf bq)\,\nonumber \\ 
&\neq& I_V (\genf a0 \W \genf bq )\,.
\label{genv.nonlinear}
\end{eqnarray}

The generalized interior product, when restricted to the ordinary
subspace of vector fields, i.e., to generalized vector fields of
the type $V = (v, 0)\,,$ is identical to the interior product of
ordinary differential geometry, acting independently on the two
ordinary forms which make up a generalized form. However, several
relations which hold for the ordinary interior product are not
satisfied by the generalized one; we list a few here.
\begin{enumerate}
\item  
The generalized interior product does not anticommute; for two
generalized vector fields $V=(v_{1},v_{0})$ and $W=(w_1,w_0)$ and
any generalized $p$-form $\genf ap = (\alpha_p, \alpha_{p+1})\,,$
\begin{eqnarray}
(I_V I_W + I_W I_V) \genf a p =
(-1)^{p-1}(w_0 i_{v_1}+v_0 i_{w_1})(0, \alpha_p)\,
\neq 0\,.
\label{genv.anticom}
\end{eqnarray}
\item Contraction of the generalized zero-form does not vanish, 
\begin{equation}
I_V \genf a0 = (0, i_{v_1}\alpha_1)\,.
\label{genv.zerocont}
\end{equation}
\item Contraction of a generalized one-form is a generalized
scalar, 
\begin{equation}
I_V \genf{a}{1} =
(i_{v_1}\alpha_1,v_0\alpha_1+i_{v_1}\alpha_2)\,. 
\label{genv.onecont}
\end{equation}
It follows that the space of generalized vector fields ${\cal
X}_G(M)$ is not the dual space of generalized $1$-form fields
$\Omega^{1}_G(M)$.
\end{enumerate}
%

%%%%%%%%%%%%%%%%%%%%%%%%%
\section{\label{lieform}Lie derivative: first attempt}
%%%%%%%%%%%%%%%%%%%%%%%%%
Equipped with the generalized exterior derivative and generalized
inner derivative we can define a generalization of the Lie
derivative. In his book~\cite{Cartanbook}, Cartan introduces a
combination of the exterior derivative and the interior product
which was coined as the ``Lie derivative" by Sledbodzinsky. Marsden
calls this Cartan's ``magic formula''.  The Lie derivative with
respect to a vector field $v$ acts on exterior differential forms
according to Cartan's magic formula:
\begin{equation}
L_v\alpha = i_v\D \alpha + \D(i_v \alpha).
%\label{}
\end{equation}
This formula does not depend on the metric imposed upon the base
space of independent variables, and has been called the homotopy
formula by Arnold.

We will take as our starting point Cartan's formula for the Lie
derivative of a $p$-form with respect to a vector, and generalize
that. However in \S\ref{lievector} we will find that the
resulting derivative is problematic when applied on a generalized
vector field and we have to add an extra correction term.

For the moment, let us define the generalized Lie derivative
$\mathcal L_V$ with respect to the generalized vector field $V$,
$\mathcal L_V: \Omega^p_G(M)\rightarrow \Omega^p_G(M)\,,$ as
\begin{equation}
\mathcal L_V\genf a p = I_V\gend\genf a p + 
\gend I_V\genf a p\,,
\label{lieform.def}
\end{equation}
which is a generalization of Cartan's formula.  Using
Eqs.~(\ref{genf.extd}) and (\ref{genv.contract}), we find that
\begin{eqnarray}
\mathcal L_V\genf a p = (L_{v_1}\alpha_p - pkv_0\alpha_p\,, && 
L_{v_1}\alpha_{p+1} - (p+1)kv_0\alpha_{p+1} \, \nonumber \\ 
&& + p(-1)^{p-1}(\D v_0)\alpha_p + (-1)^p v_0 \D\alpha_p )\,,
\label{lieform.action}
\end{eqnarray}
where as usual $\genf a p\, =(\alpha_p,\alpha_{p+1})\,,
V=(v_1,v_0)\,,$ and $L_{v_1}$ is the ordinary Lie derivative with
respect to the ordinary vector field $v_1$. This formula will be
modified later on, but let us check the consequences of this
formula here.

This generalized Lie derivative is a derivation on the space of
generalized forms, satisfying Leibniz rule and linearity,
\begin{eqnarray}
&&\mathcal L_V(\genf a p \W \genf b q ) = 
(\mathcal L_V \genf a p ) \W \genf b q +
\genf a p \W (\mathcal L_V\genf b q)\,,
\label{lieform.Leibnitz}\\
&&\mathcal L_{\lambda V+ W}=\lambda \mathcal L_V + 
 \mathcal L_W \,,
\label{lieform.linear}
\end{eqnarray}
where $V, W$ are generalized vector fields and $\lambda$ is an
arbitrary constant. 

%%%%%%%%%%%%%%%%%%%%%%%%%
\subsection{Generalized commutator and Jacobi identity}
%%%%%%%%%%%%%%%%%%%%%%%%%
We recall that the ordinary Lie derivatives $L_{v_1}$ and
$L_{w_1}$ with respect to ordinary vector fields $v_1$ and $w_1$
acting on an ordinary $p$-form $\alpha_p$ satisfy
\begin{equation}
L_{v_1}L_{w_1}\alpha_p - L_{w_1}L_{v_1}
\alpha_p=L_{[v_1,w_1]}\alpha_p\,,
\label{lieform.ordcomm}
\end{equation}
where $[v_1\,,w_1]$ is the usual commutator between the
ordinary vector fields $v_1$ and $w_1$.

We will generalize this formula to compute the generalized
commutator of two generalized vector fields. We will compute the
generalized bracket 
\begin{equation}
[\mathcal L_V\,,\mathcal L_W]\genf a p\, := (\mathcal L_V\mathcal
L_W - \mathcal L_W \mathcal L_V)\genf a p \,,
\label{lieform.bracket}
\end{equation}
and see if we can find a vector field $\{V,W\}$ such that
$[\mathcal L_V\,, \mathcal L_W ] = \mathcal L_{\{V,W\}}$.  We will
then define $\{V,W\}$ as generalized commutator of generalized
vector fields $V,W \in {\cal X}_{G}(M)$.

Expanding the right hand side of Eq.~(\ref{lieform.bracket}), we
find that it does have the form of a generalized Lie derivative of
$\genf a p$ with respect to a generalized vector field. So we can
in fact write
\begin{equation}
\mathcal L_V\mathcal L_W \genf a p - 
\mathcal L_W \mathcal L_V\genf a p = 
\mathcal L_{\{V,W\}}\genf a p\,, 
\label{lieform.bracket2}
\end{equation}
where $\{V,W\}$ is a generalized vector field, the generalized
commutator of $V$ and $W$. For $V = (v_1, v_0)\,, W = (w_1,
w_0)\,,$ the generalized commutator is calculated directly using
Eq.~(\ref{lieform.action}) to be 
\begin{eqnarray} 
\{V,W\} = \Big([v_1\,, w_1]\,, L_{v_1}w_0 - L_{w_1}v_0 \Big)
\,. 
\label{lieform.gencomm}
\end{eqnarray}
We get back ordinary commutation relation for ordinary vector
fields by setting $v_0 ,w_0 = 0$. This generalized commutation
relation can be identified with the Lie algebra of a semidirect
product ${\cal X}(M)\ltimes C^\infty(M)\,.$

Generally speaking, let $(v_1, w_1)$ be the elements of the Lie
algebra of vector fields on $M\,,$ and $(v_0, w_0)$ the elements of
$C^\infty(M)\,,$ which is an Abelian Lie algebra under
addition. The Lie algebra of the semidirect product is defined by
Eq.~(\ref{lieform.gencomm})\,.

We can check from its definition that $\{V,W\}$ is antisymmetric in
$V$ and $W$ and bilinear. And we can also calculate directly that
Jacobi identity is satisfied, for $U,V,W\in {\cal X}_G(M)\,,$
\begin{equation}
\{U,\{V,W\}\}+\{V,\{W,U\}\}+\{W,\{U,V\}\}=0\,.
\label{lieform.Jacobi}
\end{equation}
Therefore the space ${\cal X}_G(M)$ of generalized vector fields
together with the generalized commutator \{\;,\;\} form a Lie
algebra.

%%%%%%%%%%%%%%%%%%%%%%%%%
\section{\label{lievector}Lie derivative: improvement}  
%%%%%%%%%%%%%%%%%%%%%%%%%
While we could use Cartan's formula for a definition of the Lie
derivative on forms, we have to find another definition for the Lie
derivative of a vector field. We will do this by assuming that the
generalized Lie derivative is a derivation on generalized vector
fields. 

In other words, we want that the following equality should hold for
any two generalized vector fields $V, W\,,$ and any generalized
$p$-form $\genf a p\,$:
\begin{eqnarray}
\mathcal L_V (I_W \genf a p) = I_W(\mathcal L_V\genf a p) + 
I_{\mathcal L_V\,W}\genf a p \,,
\label{lievector.Leibnitz}
\end{eqnarray}
where we have written $\mathcal L_V W$ for the action of $\mathcal
L_V$ on $W$. This is what we would like to define as the Lie
derivative of $W$ with respect to $V$.  However, from
Eqs.~(\ref{genv.contract}) and (\ref{lieform.action}) we find that
\begin{equation}
\mathcal L_V (I_W \genf a p) - I_W \mathcal L_V \genf a p =
I_{([v_1,w_1] + k v_0 w_1, L_{v_1}w_0 - L_{w_1}v_0)} \genf a p - 
(-1)^p(0, L_{v_0w_1}\alpha_{p})\,.
\label{lievector.failure}
\end{equation}

We see that $\mathcal L_V$ on generalized vectors cannot be defined
if the Lie derivative of a generalized $p$-form is as in
Eq.~(\ref{lieform.action}). We can try to resolve this problem by
modifying the formula for the Lie derivative of a generalized
$p$-form, by adding an extra term of the form $(0,\beta_{p+1})$ to
the right hand side of Eq.~(\ref{lieform.action}). This ordinary
$p+1$-form $\beta_{p+1}$ must be constructed form $v_{0}$ and
$\alpha_p$, such that setting $v_0=0$ implies $\beta_{p+1}=0\,.$
Also, this $\beta_{p+1}$ should not depend on $\alpha_{p+1}$ or
$v_1$ since the extra term in Eq.~(\ref{lievector.failure}) does
not depend on these objects.  Then most general expression for
$\beta_{p+1}$ is
\begin{equation} 
\beta_{p+1}=(-1)^p(\gamma(p)v_0
\D\alpha_p+\delta(p)(\D v_0)\alpha_p)\,.
\label{lievector.extra}
\end{equation}
With this extra term, we can define the modified Lie derivative
$\hat{\mathcal L}_V$ as
\begin{eqnarray}
\hat{\mathcal L}_V \genf ap = {\mathcal L}_V \genf ap + (0,
\beta_{p+1})\,.
\label{lievector.hatlie}
\end{eqnarray}
Now if we calculate ${\hat {\cal L}_V} I_W \genf a p - I_W {\hat
{\cal L}_V} \genf a p\,,$ we find that we must have $\delta(p)=p$
and $\gamma(p)=-1$ in order that ${\hat {\cal L}_V}W$ is
well-defined.  Therefore we get
\begin{eqnarray}
{\hat {\cal L}_V}\genf ap &=& \mathcal L_V\genf ap +
(-1)^p(0, -v_0 \D\alpha_p + p \D v_0 \alpha_p) \, \nonumber \\ 
&=& (L_{v_1}\alpha_{p}-pk v_0\alpha_{p}, L_{v_1}\alpha_{p+1} - 
(p+1)kv_0 \alpha_{p+1})\,.
\label{lievector.lieform} 
\end{eqnarray}
This is a modification of the formula given in
Eq.~(\ref{lieform.action})\,. 

It can be checked easily using Eq.~(\ref{lievector.lieform}) that
this new and improved generalized Lie derivative satisfies the
Leibniz rule,
\begin{equation}
{\hat {\cal L}_V}(\genf ap \W \genf bq) = ({\hat {\cal L}_V}\genf 
ap) \W \genf bq + \genf ap \W ({\hat {\cal L}_V} \genf bq)\,,
\label{lievector.leibniz}
\end{equation}
where $V\in {\cal X}_{G}, \genf ap \in \Omega^p_G$ and
$\genf bq\in \Omega^q_G$.

We can now write the generalized Lie derivative of a generalized
vector field. Using Eq.~(\ref{genv.contract}) and
(\ref{lievector.lieform}) we find
\begin{equation}
{\hat {\cal L}_{V}} I_W - I_W {\hat {\cal L} _{V}} 
= I_{([v_1,w_1] + kv_0w_1, L_{v_1}w_0)}\,.
\label{lievector.derivation}
\end{equation}
Therefore we can define the generalized Lie derivative of a
generalized vector field as
\begin{equation}
{\hat {\cal L}_V}W = ([v_1,w_1] + k v_0 w_1, L_{v_1}w_0)\,,
\label{lievector.lievector} 
\end{equation}
where $V=(v_1, v_0), W = (w_1, w_0)\,,$ and $ v_1,w_1\in {\cal
X}(M)\,,$ $v_0,w_0 \in C^{\infty}(M)$.

Note that $v_0, w_0 = 0$ gives the ordinary Lie derivative of an
ordinary vector field. Also, the new generalized Lie derivative on
generalized vector fields is not the generalized commutator.  It
can be checked using the new and improved definition that the
commutator of two generalized Lie derivatives is also a generalized
Lie derivative as in Eq.~(\ref{lieform.bracket2}) and definition of
generalized commutator of two generalized vectors remains the same
as in Eq.~(\ref{lieform.gencomm}),
\begin{equation}
{\hat {\cal L}_V}{\hat {\cal L}_W} - {\hat {\cal L}_W}{\hat {\cal
L}_V}   = {\hat {\cal L}_{\{V,W\}}}\,. \qquad V,W \in {\cal
X}_{G}\, 
\label{lievector.comm}
\end{equation}
Note that both sides of this equation can be taken to act either on
a generalized $p$-form or on a generalized vector.

%%%%%%%%%%%%%%%%%%%%%%%%%
\section{\label{altvect}Alternative definition of a generalized
vector field}
%%%%%%%%%%%%%%%%%%%%%%%%%
In \S\ref{genf} we defined a generalized $p$-form as an ordered
pair of an ordinary $p$-form and an ordinary $p+1$-form, and
equipped the space of these objects with a generalized wedge
product and a generalized exterior derivative. All calculations
were done using this definition. As was mentioned there, we could
have used an alternative definition, assuming the existence of a
``$-1$- form'' $\zeta$ such that $d\zeta = k$ where $k$ is a
constant. Then a generalized $p$-form can be defined as $\genf ap =
\alpha_p + \alpha_{p+1}\zeta\,,$ 
%%% %
%%% \begin{equation}
%%% \genf ap = \alpha_p + \alpha_{p+1}\zeta\,,
%%% %\label{}
%%% \end{equation}
%%% %
and has identical properties with the generalized $p$-form defined
as an ordered pair.

Coming back to generalized vector fields, we defined these objects
as ordered pairs in \S\ref{genv} and did all the calculations using
this definition. The purpose of this section will be to give an
alternative definition of the generalized vector field in analogy
with the $-1$-form of Sparling et al. and to show that the results
found so far follow from using this definition also. Let us then
define a generalized vector field as the sum of an ordinary vector
field and a scalar multiple of a special vector field $\bar X\,,$
\begin{equation}
V=v_1 + v_0 \bar X\,.
\label{altvect.vectdef}
\end{equation}
The vector field $\bar X$ is not an ordinary vector field, but may
be thought of as the unit generalized vector. It will be defined by
its action on ordinary $p$-forms,
\begin{equation}
i_{\bar X}\alpha_p=p(-1)^{p-1}\alpha_p\W\zeta\,,
\label{altvect.special}
\end{equation}
for any $\alpha_p\,\in\Omega^p(M).$ Note that this formula for
$i_{\bar X}$ automatically satisfies the Leibniz rule for
contractions,
\begin{equation}
i_{\bar X}(\alpha_p \beta_q) = (i_{\bar X}\alpha_p) 
\beta_q + (-1)^p\alpha_p (i_{\bar X}\beta_q)\,.
\label{altvect.leibniz}
\end{equation}
Note also that $i_{\bar X}\genf a p = i_{\bar X}\alpha_p$, where
as usual, $ \genf a p = \alpha_p + \alpha_{p+1} \zeta\,.$

We can define the contraction of a generalized $p$-form by an
ordinary vector field $v_1\in {\cal X}(M)$ as~\cite{nurob:cqg02}
\begin{equation}
i_{v_1}\genf ap = i_{v_1}\alpha_p +
(i_{v_1}\alpha_{p+1})\W\zeta\,. 
\label{altvect.ordcont}
\end{equation}
Using the definitions given in Eqs.(\ref{altvect.special}) and
(\ref{altvect.ordcont}) we can now define the contraction of a
generalized form by a generalized vector field $V = v_1 + v_0 \bar
X$ as
\begin{equation}
i_V\genf a p = i_{v_1}\alpha_p + \left(p(-1)^{p-1}v_0
\alpha_p +i_{v_1}\alpha_{p+1}\right)\W\zeta\,.
\label{altvect.corrcont}
\end{equation}
Clearly, this agrees with the contraction formula defined earlier
in Eq.~(\ref{genv.contract})\,.

We can use this contraction formula in conjunction with Cartan's
magic formula to calculate the Lie derivative of a generalized form
with respect to a generalized vector field. Let us first calculate
the Lie derivative of a generalized form with respect to an
ordinary vector field $v_1\,.$ Using Cartan's formula we can write  
\begin{eqnarray}
L_{v_1}\genf ap = L_{v_1}\alpha_p + (L_{v_1}\alpha_{p+1})\W\zeta
\,. 
%\label{}
\end{eqnarray}
We can now calculate the Lie derivative of $v_0\bar X$ with respect
to an ordinary vector field $w_1$. Using the results obtained so
far, we get,  for any generalized $p$-form $\genf a p\,,$
\begin{eqnarray}
L_{w_1}i_{v_0\bar X}\genf a p -  
i_{v_0\bar X} L_{w_1}\genf a p\, &=& p(-1)^{p+1}(L_{w_1}v_0)\,
\alpha_p\W\zeta \nonumber \\
&=& i_{(L_{w_1}v_0)\bar X}\genf ap  \,. 
\label{altvect.ordlie}
\end{eqnarray}
Therefore, 
\begin{equation}
L_{w_1}(v_0 \bar X) = (L_{w_1}v_0)\bar X\,, \qquad L_{w_1}\bar X =
0\,. 
\label{altvect.liexbar}
\end{equation}

The Lie derivative of a generalized form with respect to the
special vector field $\bar X$ can be calculated using Cartan's
formula, 
\begin{eqnarray}
{\cal L}_{v_0 \bar X} \genf ap = - pkv_0\alpha_p\,
&+& \Big( - (p+1)kv_0\alpha_{p+1}
 \nonumber \\ && \qquad  + p(-1)^{p-1}(\D v_0)\alpha_p 
+ (-1)^p v_0 \D\alpha_p \Big) \W\zeta\,.
%\label{}
\end{eqnarray}
Finally, combining the results obtained so far, we get the Lie
derivative of a generalized form $\genf ap = \alpha_p +
\alpha_{p+1}\W\zeta$ with respect to a generalized vector $V = v_1
+ v_0\bar X$,
\begin{eqnarray}
{\cal L}_V \genf ap &\equiv& (\gend\, i_V + i_V \gend)\genf ap\,
\nonumber \\
&=& L_{v_1}\alpha_p - pkv_0\alpha_p\, 
+ \Big(L_{v_1}\alpha_{p+1} - (p+1)kv_0\alpha_{p+1} \nonumber \\ 
&& \qquad \qquad + p(-1)^{p-1}(\D v_0)\alpha_p + (-1)^p v_0
\D\alpha_p \Big) \W\zeta\,.
%\label{}
\end{eqnarray}
Let us now try to define the Lie derivative of an ordinary vector
field $w_1$ with respect to $v_0\bar X$ can be defined. It is
straightforward to calculate that
\begin{equation}
\mathcal L_{v_0\bar X}i_{w_1}\genf ap -i_{w_1}\mathcal L_{v_0\bar
X} \genf ap  = i_{k v_0 w_1 - (L_{w_1}v_0)\bar X} \genf a p - 
(-1)^p(L_{v_0w_1}\alpha_p)\W\zeta\,.
%\label{}
\end{equation}

Because of the last term, the Lie derivative of an ordinary vector
with respect to a generalized vector cannot be defined. However, we
can correct the situation by modifying the Lie derivative,
following a procedure similar to that used for calculating the
correction term in \S\ref{lievector}. Then we find that the
Lie derivative with respect to the special vector field $v_0\bar X$
needs to be defined as
\begin{eqnarray}
\lie_{v_0\bar X}\genf a p &=& \mathcal L_{v_0\bar X} \genf a p +
(-1)^{p}(- v_0 \D\alpha_p + p (\D v_0) \alpha_p)\W\zeta\, \nonumber
\\
&=&  - pkv_0\alpha_p - (p+1)kv_0\alpha_{p+1}\W\zeta \,.
\label{altvect.liecorr}
\end{eqnarray}
It follows from this definition that
\begin{equation}
\lie_{v_0\bar X}i_{w_1} \genf ap - i_{w_1} \lie_{v_0\bar X} \genf
ap \, = 
i_{kv_0 w_1} \genf ap \,,
\label{altvect.liecontract}
\end{equation}
from which we can write
\begin{equation}
\lie_{v_0\bar X}w_1 = k v_0 w_1\,.
\label{altvect.ordvect}
\end{equation}
Also, we can calculate from this definition that
\begin{eqnarray}
\lie_{v_0\bar X} i_{w_0\bar X} \genf ap -  i_{w_0\bar
X}\lie_{v_0\bar X} \genf ap &=& 0\,, \\
\Rightarrow \qquad \lie_{v_0\bar X} {w_0\bar X} &=& 0\,.
\label{altvect.liexx}
\end{eqnarray}
So the complete expression for the modified Lie derivative, acting
on generalized forms, is
\begin{eqnarray}
\lie_V \genf ap &=& (L_{v_1}\alpha_{p}-pk v_0\alpha_{p},
L_{v_1}\alpha_{p+1} -  (p+1)kv_0 \alpha_{p+1})\,.
\label{altvect.lieform}
\end{eqnarray}
This Lie derivative also satisfies the Leibniz rule on wedge
products of generalized forms.

Now we can directly calculate the generalized Lie derivative of a
generalized vector field as
\begin{equation}
\lie_{v_1 +v_0\bar X}(w_1 + w_0 \bar X)=[v_1,w_1] + k v_0 w_1 + 
L_{v_1}w_0 \bar X\,.
\label{altvect.lievector}
\end{equation}
We can also calculate the commutator of Lie derivatives to find
that 
\begin{equation}
\lie_{(v_{1}+v_{0}\bar X)} \lie_{(w_{1}+w_{0}\bar X)} - 
\lie_{(w_{1}+w_{0}\bar X)} \lie_{(v_{1}+v_{0}\bar X)} = 
\lie_{([v_{1},w_{1}]+\bar X(L_{v_{1}}w_{0}-L_{w_{1}}v_{0}))}\,. 
\label{altvect.liecomm}
\end{equation}
So the generalized commutator can be defined as 
\begin{equation}
[V,W]=[v_1, w_1]+\bar X(L_{v_1}w_0 - L_{w_1}v_0)\,,
%\label{}
\end{equation}
where $V = v_1 + v_0\bar X, \; W=w_1 + w_0\bar X\,.$

This generalized commutator satisfies Jacobi identity, as can be
checked easily. Hence, we obtain all the earlier relations of
\S\ref{genv}, \S\ref{lieform} and \S\ref{lievector} using the
alternative definition of generalized vector fields. 

%%% Note that we can calculate generalized commutator directly  
%%% considering $\bar X$ commutes with ordinary vector fields.  

%%%%%%%%%%%%%%%%%%%%%%%%%
\section{\label{relpart}Application: Relativistic free particle}
%%%%%%%%%%%%%%%%%%%%%%%%%
Carlo Rovelli analysed the Hamiltonian formalism of the
relativistic mechanics of a particle~\cite{Rovelli:2002ef}. This
analysis can be formulated in the language of generalized vector
fields and generalized forms.

Let a system be described by Hamiltonian $H_0(t,q^i,p_i)$,
where $q^{i}$ are the coordinates, $p_{i}$ are the conjugate
momenta and $t$ is the time. Now we can define an extended
coordinate system as $q^a=(t,q^i)$ and corresponding momenta as
$p_{a}=(\Pi,p_{i})$. We define a curve in this extended space as
$m:\tau \rightarrow (q^{a}(\tau),p_{i}(\tau))$. Here the
constrained Hamiltonian will be
\begin{equation}
H(q^a,p_a)= \Pi + H_0(t,q^i, p_i)=0\,.
\label{relpart.ham}
\end{equation}
Now if we consider the Hamiltonian equation for this constrained
Hamiltonian, following Rovelli we can write
\begin{equation}
I_V \Omega = 0\,,
\label{relpart.dyn}
\end{equation}
where $V$ is a generalized vector, $V=(v_1, v_0)\,,$ and $\Omega$
is a generalized 2-form. Let us also define $\Omega$ as
\begin{equation}
\Omega=(\omega,\omega \theta)\,.
\label{relpart.symp}
\end{equation}
Here $\theta=p_a \D q^a$ is the canonical one-form and
$\omega=\D\theta$. So we get 
\begin{eqnarray}
\hfill i_{v_1}\omega &=& 0\,, 
\label{relpart.sympdyn}\\
-2v_{0}\omega + i_{v_{1}}(\omega \theta) &=& 0\,. 
\label{relpart.constraint}
\end{eqnarray}
We can identify Eq.~(\ref{relpart.sympdyn}) as the same as the one
considered by Rovelli and so $v_1$ is the tangent vector of the
curve $m$
\begin{equation}
v_{1}=\dot{q^a}\frac {\partial}{\partial q^{a}}+\dot{p_{i}}\frac
{\partial}{\partial p_{i}}\,,
\label{relpart.tangent} 
\end{equation}
where a dot denotes the differentiation with respect to $\tau$. Now
using Eq.s~(\ref{relpart.sympdyn}), (\ref{relpart.constraint}) and
(\ref{relpart.tangent}), we find
\begin{eqnarray}
v_{0} &=& 1/2(p_{i}\dot{q^{i}}-H_{0}\dot{t}) \nonumber \\
&=&1/2(p_{i}\dot{q^{i}}+\Pi \dot{t})\,.
\label{relpart.lag}
\end{eqnarray}
The function within the bracket in right hand side is exactly the
constrained Lagrangian. So $v_0$ is propotional to the
constrained Lagrangian of the system.

\bigskip
{\bf References}

%%%%%%%%%%%%%%%%%%%%%%%%%%%%%%%%%%%%%%%%%%%%%%%%%%


\begin{thebibliography}{99}
%%% \frenchspacing
%\baselineskip=23pt

\bibitem{Sparling:avrc97}
G~A~J~Sparling
abstract/virtual/reality/complexity {\sl in Geometry and Physics}
Eds. L~Mason and K~P~Tod, Oxford University Press, 1997

\bibitem{Perjes:esi98}
Z~Perj\'es and G~A~J~Sparling 
The Abstract Twistor Space
of the Schwarzschild Space-Time {\sl Preprint} ESI Vienna no. 520

\bibitem{nurob:cqgl01}
P~Nurowski and D~C~Robinson {\sl Class.~Quant.~Grav.} {\bf 18}
(2001) L81.

\bibitem{nurob:cqg02}
P~Nurowski and D~C~Robinson {\sl Class.~Quant.~Grav.} {\bf 19}
(2002) 2425

%\cite{Guo:2002yc}
\bibitem{Guo:2002yc}
  H.~Y.~Guo, Y.~Ling, R.~S.~Tung and Y.~Z.~Zhang,
  % ``Chern-Simons term for BF theory and gravity as a generalized
  % topological   %field theory in four dimensions,''
  Phys.\ Rev.\ D {\bf 66} (2002) 064017
%  [arXiv:hep-th/0204059].
  %%CITATION = HEP-TH 0204059;%%



\bibitem{Robinson:jmp03}
D~C~Robinson {J.~Math.~Phys.} {\bf 44} (2003) 2094

\bibitem{Cartanbook}
E.~Cartan {\sl Lecon sur la th\'eorie des espaces \`a connexion
projective} Gauthier-Villars, Paris, 1937

%\cite{Rovelli:2002ef}
\bibitem{Rovelli:2002ef}
  C.~Rovelli
  Covariant Hamiltonian formalism for field theory: Symplectic
  structure  and 
  Hamilton-Jacobi equation on the space G 
  {\sl preprint} arXiv:gr-qc/0207043
  %%CITATION = GR-QC 0207043;%%




\end{thebibliography}
\end{document}